\documentclass{amsart} 
\usepackage{graphicx}
\usepackage{amstext}

\newcommand{\be}{\begin{equation}}
\newcommand{\ee}[1]{\label{#1} \end{equation}}
\newcommand{\ba}{\begin{eqnarray}}
\newcommand{\ea}[1]{\label{#1} \end{eqnarray}}
\newcommand{\nl}{\nonumber \\}
\newcommand{\re}[1]{(\ref{#1})}

\newcommand{\ave}{\bar{u}}
\newcommand{\pd}[2]{ \frac{\partial #1}{\partial #2}}

\begin{document}

\title{About the temperature of moving bodies }
\author{Tam\'as S. B\'ir\'o$^{1}$ and P\'eter V\'an$^{1,2}$}
\address{
KFKI Research Institute for Particle and Nuclear Physics, Budapest\\
BME, Department of Energy Engineering, Budapest 
}


\begin{abstract}
Relativistic thermodynamics is constructed from the point of view of special 
relativistic hydrodynamics. A relativistic four-current for heat and a general
treatment of thermal equilibrium between moving bodies is presented. 
The different temperature transformation formulas of Planck
and Einstein, Ott, Landsberg and Doppler appear upon particular assumptions
about internal heat current.
\end{abstract}

\maketitle

\section{Introduction}
Considering the temperature of moving bodies, the easier question is to answer, 
what is the apparent spectral temperature. In this case a spectral parameter is 
transformed if the thermalized source is moving with respect to the observer 
(detector system), and the transformation rule can be derived from that of 
the energy and momentum in the co-moving system. 

This has been known from the beginnings of the theory of special relativity and 
never has been seriously challenged. An essentially tougher problem is to 
understand the relativistic thermalization: what is the intensive 
parameter governing the state with energy exchange equilibrium between two, 
relatively moving bodies in the framework of special relativity. In particular 
how this general temperature should transform and how does it depend on the speed 
of the motion. Here several answers has been  historically offered, practically 
including all possibilities.

Planck and Einstein concluded that moving bodies are cooler by a Lorentz factor 
\cite{Pla07a,Ein07a,Pla08a}, first Blanu\u sa then Ott has challenged 
this opinion \cite{Bla47a,Ott63a} by stating that on the contrary, such bodies 
are hotter by a Lorentz factor. During later disputes several authors supported
one or the other view  (see e.g. \cite{EbeKuj67a,TerWer71a,Tre77a,Mol72b,New80a,Liu96a,
Req08m,Sew08a} and the references therein) and also some new opinions 
emerged. Landsberg argued for unchanged values of the temperature \cite{Lan66a,Lan67a}. 
Other authors observed that for a thermometer in equilibrium with black body radiation 
the temperature  transformation is related to the Doppler formula 
\cite{CosMat95a,LanMat96a,LanMat04a,CasJou03a,MiAta09a}, therefore the measured temperature 
seems to  depend on {the physical state of} the thermometer. This problem is circumvented by the suggestion
that thermal equilibrium  would have a meaning only in case of equal velocities
\cite{vKa68a,Lan78b,Eim75a}.  

Behind these different conclusions there are, {in our opinion}, different views about 
the energy transfer and mechanical work, and the identification of the heat
{\cite{Req08m}}. In a simplifying manner the assumptions 
and views about the Lorentz transformation properties of internal energy, work, 
heat, and entropy influence such properties and the very definition of the 
absolute temperature. Coming to the era of fast computers, a renewed interest 
emerged in such questions by modelling stochastic phenomena at relativistic 
energy exchanges and relative speeds \cite{CubAta07a,DunHan09a,MonAta09a}. 
In particular, dissipative hydrodynamics applied to high energy heavy ion collisions 
requires the proper identification of temperature and entropy 
\cite{Mur04a,KoiAta07a,DenAta09a,BaiRom07a,OsaWil08a,DumAta07a,MolHou08a,Mol09a,SonHei09a}. 
In this letter we show that our approach to replace the Israel-Stewart theory
of dissipative hydrodynamics, proposed earlier \cite{VanBir08a,BirAta08a,Van08a,Van09a},
is related to the problem of thermalization of relatively moving bodies with 
relativistic velocities and our suggestion is compatible with the 
foundations of thermodynamics and guarantees causal heat propagation.

o\o

By doing so we encounter the following questions in our analysis:
\begin{enumerate}
\item   What moves (or flows)? Total energy and momentum do flow correlated, 
but further conserved charges (baryon number, electric charge, etc.) may flow 
differently. In relativistic systems one has to deal with the possibility 
that the velocity field is not fixed to either current, not being restricted 
to the Landau-Lifshitz \cite{LanLif59b} or Eckart \cite{Eck40a3} frames.
\item   What is a body? We exploit, how do integrals over extended volumes 
relate to the local theory of hyd\-rodynamics, and what is a good local definition 
for volume change in relativistic fluids. In close relation to this, we suggest
a four-vector generalization to the concept of heat.
\item   What is a proper equation of state? Here the functional dependency between entropy and the relativistic internal energy is fixed to a
particular form. \item What is the proper transformation of the temperature? As we have mentioned 
above prominent physicists expressed divergent opinions on this in the past. 
This problem is intimately related to that of thermal equilibrium and to the 
proper description of internal energy.
\end{enumerate} 

\section{Hydro- and thermodynamics} In this letter we concentrate on the energy-momentum density of a
one-component fluid,
but the results can be generalized considering conserved currents in multi-component systems easily. The energy-momentum tensor can be split
into components 
aligned to the fiducial four-velocity field, $u^a(x)$, and orthogonal ones:
\ba
 T^{ab} &=& eu^au^b + u^aq^b + q^au^b + P^{ab}
\ea{SPLIT}
with $u_aq^a=0$ and $u_aP^{ab}=P^{ab}u_b=0$. When considering complex 
systems, like a quark-gluon plasma, 
the velocity field can be aligned only with one of the conserved currents, 
unless several currents 
are parallel (i.e. different conserved charges are fixed to the same carriers). 
In our present treatment the velocity field is general. 

Relativistic thermodynamics is obtained by integrating the local energy-momentum 
conservation
on a suitably defined extended and homogeneous thermodynamic body.
  Therefore in the balance of energy-momentum we separate the terms perpendicular
and parallel to the velocity field    as
\ba
 \partial_bT^{ab} &=& \frac{d}{d\tau} ({e}u^a+{q}^a) +(eu^a+q^a) \partial_b u^b \nl
      &+& p(\frac{d}{d\tau} u^a+ u^a\partial_bu^b) - \frac{du_b}{d\tau} (u^a q^b+\Pi^{ab}) \nl
     &- & \nabla^a p + \nabla_b(u^a q^b + \Pi^{ab}) = 0^{a}.\ \ \
\ea{ENERGIZE}

From now on the 
proper time derivative is denoted by a dot \hbox{$\dot f = df/d\tau = u_a\partial^a f$} 
for an arbitrary function $f(x)$. 
$\nabla^a=\partial^a-u^au_c\partial^c$
denotes a derivative perpendicular to the velocity field and we also split 
the pressure tensor into a hydrostatic part and 
a rest: $P^{ab}=p(u^au^b-g^{ab})+\Pi^{ab}$.

Let us now assume, that
$u^a$ is smooth and we may give a connected smooth surface $H$  that is initially
perpendicular 
to the velocity field and has a smooth (two -dimensional) boundary. As a further simplification we will 
assume that the velocity field is not accelerating $\dot u^a = 0^a$, 
therefore $\partial_au^a=\nabla_au^a$ and the hypersurface 
 remains 
perpendicular to the four-velocity field. Hence the propagation of the surface can be characterized 
by the proper time $\tau$ of any of its wordlines. We refer to this hypersurface - 
a three dimensional spacelike set related to our fluid - as a  
\textit{thermodynamic body}. 
Considering homogeneous  bodies we set $\nabla_a e =0$ and  $\nabla_a p =0$.
It is important that the velocity field itself is not homogeneous, $\nabla^au^b\ne 0$.  Now integration of \re{ENERGIZE} on $H(\tau)$ results in
\ba
\int_{H(\tau)}(\dot{e}u^a+\dot{q}^a +(eu^a+q^a) \partial_b u^b 
     +p u^a\partial_bu^b)dV\!=\!\!\nonumber\\
\int_{H(\tau)}\nabla_b(u^a q^b\!+\Pi^{ab})dV.
\ea{no} 
 With the above conditions 
wa apply the transport theorem of Reynolds to the l.h.s. of eq.(\ref{no}) and the Gauss-Ostrogradsky theorem
to the r.h.s. of eq.(\ref{no}) and obtain
\be
\dot{E}\bar u^a+\dot{G}^a +p \bar u^a\dot V=
\oint_{\partial H(\tau)}\limits\!\!\!\left(u^a q^b\!+\Pi^{ab}\right) \,
  dA_b = \delta Q^a.
\ee{hom} 
Here $\ave^a = \int_Hu^a dV/V$ is the average velocity field inside $H$,
$E=eV$ is the total energy, $G^a = \int_Hq^a dV,$ and $dA_b$
is the two-form surface measure circumventing the homogeneous body in the region
$H(\tau)$. The two-dimensional surface integral term is the physical energy and momentum leak
(dissipation rate) from the  body under study, we denote it by $\delta Q^a$. This 
is a four-vector generalization of the concept of heat.
It describes both energy and momentum transfers to or
from the homogeneous body.

The derivation of the temperature in thermodynamics is related to the maximum 
of the total entropy of a system (under various constraints). This way its 
reciprocal, $1/T$ is an integrating factor to the heat in order to obtain 
a total differential of the entropy \cite{Cal85b,Mat05b}. Here we follow the same 
strategy considering a vectorial integrating factor $A^a$: 
\be
\delta Q^a = \dot{E^a} + p \bar u^a\dot V = A^a  \dot{S} + \Sigma^a
\ee{CLAUSIUS_FACTOR}
with ${E}^a=E\ave^a+G^a$ the energy-momentum vector of the body, $\dot{\ave}^a=0$ and
$\Sigma^a$ orthogonal to $A^a$.
The decisive point is, that -- according to the above -- the 
entropy of the homogeneous body is a function of the energy-momentum vector
and   the volume: $S=S(E^a,\text{V})$. Multiplying eq.(\ref{CLAUSIUS_FACTOR}) by $A_ad\tau/(A_bA^b)$ and utilizing that
$d\ave^a=0$ we obtain
\be
 dS = \frac{A_a\ave^a}{A_bA^b} \, dE + \frac{A_a}{A_bA^b} \, dG^a  
        + p \frac{A_a\ave^a}{A_bA^b}  \, d\text{V}.
\ee{DIFF_ENTROPY}
The connection to classical thermodynamics is best established by the temperature
definition
\be
 \frac{1}{T} := \frac{A_a\ave^a}{A_bA^b}.
\ee{TEMP_DEFINITION}
The intensive parameter associated to the change of the four-vector $G^a$ is
denoted by
\be
 \frac{g^a}{T} := \frac{A^a}{A_bA^b}.
\ee{THETA_DEFINITION}
With these notations we arrive at the following form of the Gibbs relation:
\be
 TdS = dE + g_a dG^a + pdV.
\ee{GIBBS_RELATION}
Due to the definitions eq. (\ref{TEMP_DEFINITION},\ref{THETA_DEFINITION})
$g_a\ave^a=1$. Hence the Gibbs relation can be written in
the alternative form
\be
 TdS = g_a \, dE^a + pdV,
\ee{GIBBS_NEW_FORM}
suggesting that the traditional change of the energy, $dE$, has to be generalized
to the change of the total energy-momentum four-vector, \hbox{$dE^a=d(E\ave^a+G^a).$}

For the well-known
J\"uttner distribution \cite{Jut11a} $g^a=\ave^a$.  This equality has been postulated
among others in the classical theory of Israel and Stewart \cite{IsrSte80a}.
Then \re{GIBBS_NEW_FORM}
reduces to 
\be
 TdS =  d\tilde  E + pdV,
\ee{GIBBS_CLASSICAL_FORM}
where \( \tilde E = \bar u_a E^a\). In this case the internal energy can
be interpreted as $\tilde E$, but its total differential contributes to the
Gibbs relation. In the general case  $g^a\neq\ave^a$ - considered below -
there
remains a term, related to momentum transfer. It is reasonable to assume that
the new intensive variable is timelike: \hbox{\(g_{a}g^a\geq 0\)}. Then we introduce 
\begin{equation}
w^a := g^a - \bar u^a. 
\label{gsplit}\end{equation}
Now $\|w^{a}\|^2=-w^aw_a=-g^ag_a +1 {\leq}
1$ follows.  The spacelike four-vector $w^a$ has the physical dimension of
velocity. Due to $-1\leq w^a w_a \leq 0$ and $\bar u_a w^a=0$ its general
form is given by $w^a=(\gamma v|{\mathbf w}|,\gamma {\mathbf w})$. In this
case $|{\mathbf w}|^2\leq 1$. We interpret $\mathbf w$ as the \emph{velocity of
the internal energy current}.

Here some important physical questions arise: is it only a single or several differential
terms describing the change of energy and momentum? When two, relatively moving
bodies come into thermal contact what can be exchanged among them in the evolution
towards the equilibrium? 

\section{Two bodies in equilibrium} Let us now consider two different bodies with different average velocities and energy currents.
When all components of $E^a$ and the total volume are kept constant independently, i.e. $dE^a_1+dE_2^a=0$
and $dV_1+dV_2=0$ while $dS(E_1^a,V_1)+dS(E_2^a,V_2)=0$ in the entropy maximum,
then from (\ref{GIBBS_NEW_FORM}) we obtain the conditions
\be
 \frac{g^a_1}{T_1} = \frac{g^a_2}{T_2}, \qquad \frac{p_1}{T_1} = \frac{p_2}{T_2}. 
\ee{GENERAL_THERMAL_EQUILIBRIUM}
This, in general, does not mean the equality of temperatures. 

In order to simplify the discussion we restrict ourselves to
one-dimensional motions and consider \hbox{$\ave^a=(\gamma,\gamma v)$}
with $\gamma=1/\sqrt{1-v^2}$ Lorentz-factors. The energy current velocity
is given by \hbox{$w^a=(\gamma v w, \gamma w)$} and \hbox{$q^a={(\gamma(1+v w)},{ \gamma (v+w))}$}.
Here $w$ describes the speed of internal energy current.
The thermal equilibrium condition (\ref{GENERAL_THERMAL_EQUILIBRIUM}) hence
requires
\ba
 \frac{\gamma_1(1+v_1w_1)}{T_1} &=& \frac{\gamma_2(1+v_2w_2)}{T_2},  \nl
 \frac{\gamma_1(v_1+w_1)}{T_1} &=& \frac{\gamma_2(v_2+w_2)}{T_2}.
\ea{SPECIAL_EQUIL}
The ratio of these two equations reveals that in equilibrium the composite 
relativistic velocities are equal, 
\be
 \frac{v_1+w_1}{1+v_1w_1} = \frac{v_2+w_2}{1+v_2w_2},
\ee{VELOCITY_EQUIL}
and the difference of their squares leads to
\be
  \frac{\sqrt{1-w_1^2}}{T_1} = \frac{\sqrt{1-w_2^2}}{T_2}.
\ee{SCALAR_TEMP_EQUIL}
The equality of some other velocities were investigated by
several authors \cite{vKa68a,Lan78b,Eim75a,Liu94a}.

One realizes that in the thermal equilibrium condition four velocities are involved
for a general observer: $v_1$, $v_2$, $w_1$ and $w_2$.
By a Lorentz transformation {only} one of them can be eliminated.
{ The remaining three (relative) velocities reflect physical conditions in
the system.}  According to eq.(\ref{VELOCITY_EQUIL})
\be
 w_1 = \frac{v+w_2}{1+vw_2}
\ee{VEL_EQUIL_ECKART}
with $v=(v_2-v_1)/(1-v_1v_2)$ relative velocity.
The associated factor, $\sqrt{1-w_1^2}$ can be expressed
and the  temperatures satisfy
\be
 T_1 = T_2 \, \frac{\sqrt{1-v^2}}{1+vw_2}.
\ee{ECKART_TEMP}
This {includes the} general Doppler formula \cite{CosMat95a,LanMat96a,LanMat04a,
CasJou03a,MiAta09a,Dem85b}. 

\begin{figure}
\center\includegraphics[width=0.48\textwidth]{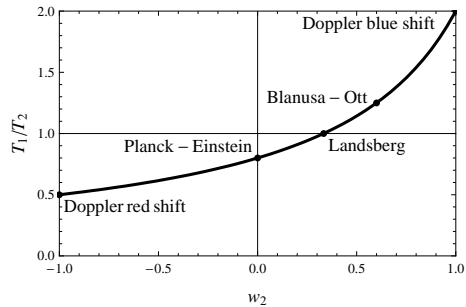}
\caption{\label{Fig1}
Ratio of the temperatures of the observed body in its rest frame, $T_2$ 
to that shown by an ideal thermometer, $T_1$ as a function of the 
the speed of the heat current in the body, $w_2$ while approaching with the 
relative velocity $v=-0.6$.
}
\end{figure}

It is enlightening to investigate this formula with different assumptions
about the energy current speed in the observed body, $w_2$.
The induced energy current speed in an ideal thermometer, $w_1$ and the
temperature it shows, $T_1$, are now determined by eqs.(\ref{VEL_EQUIL_ECKART}) and (\ref{ECKART_TEMP}).
Figure \ref{Fig1} plots temperature ratios $T_1/T_2$ for a body closing
with $v=-0.6$ as a function of the energy current speed, $w_2$.

\begin{figure}
\center\includegraphics[width=0.4\textwidth]{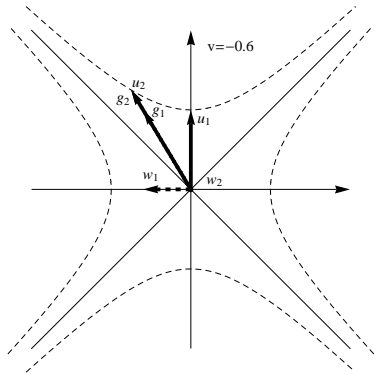}
\caption{\label{Fig2}
The space-time figure for the Planck-Einstein rule of two 
thermodynamic bodies in equilibrium. There is no  energy current in  the
observed body ($w^{a}_2=0$), therefore the $u^{a}_2$  four-velocity (solid arrow)
is parallel to the vectors ($g^a_1, g^a_2$).
The ratio of the temperatures is $T_1/T_2 <1$. 
}
\end{figure}

\begin{figure}
\center\includegraphics[width=0.4\textwidth]{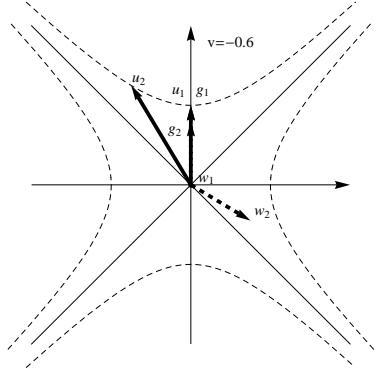}
\caption{\label{Fig3}
The space-time figure for the Blanu\v{s}a-Ott rule. The energy current stands
in the thermometer
($w^{a}_1=0$), therefore the $u^{a}_1$  four-velocity (solid arrow)
is parallel to the vectors  ($g^a_1, g^a_2$). The ratio of the temperatures is $T_1/T_2>1$.
}
\end{figure}

\begin{figure}
\center\includegraphics[width=0.4\textwidth]{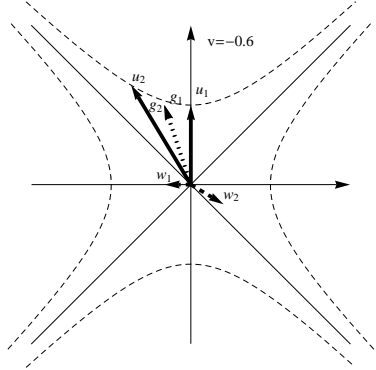}
\caption{\label{Fig4}
The space-time figure for the Landsberg rule. There is no energy current in
the composed system, therefore the four-vectors  
($g^a_1$, $g^a_2$, dotted arrows) are equal. The ratio of temperatures is 
$T_1/T_2=1$. Here $w_1=-0.33$, $w_2=0.33$.
}
\end{figure}

\begin{figure}
\center\includegraphics[width=0.4\textwidth]{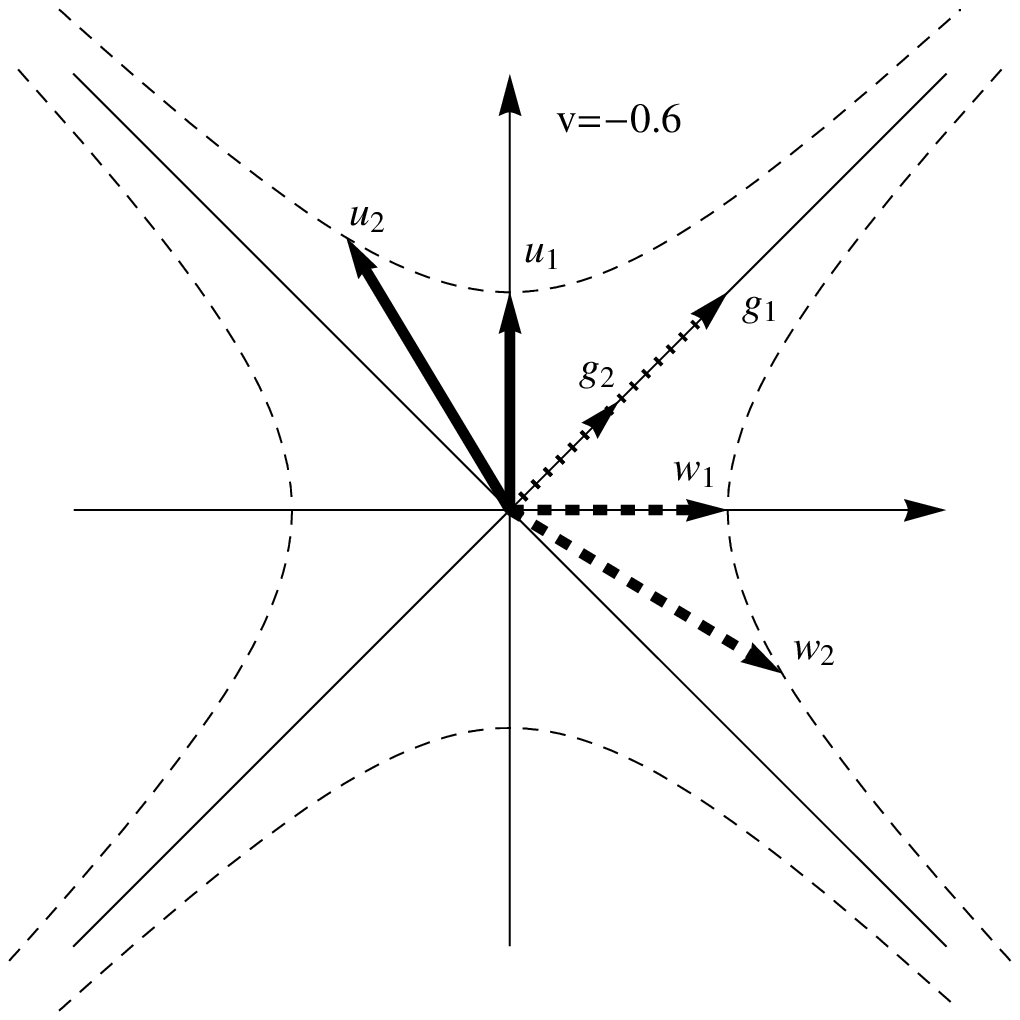}
\caption{\label{Fig5}
The space-time figure for the Doppler red shift rule of two 
thermodynamic bodies in equilibrium. The energy current speed (dashed 
arrows) in the observed body is that of the light $w_2=1$, therefore $w_1=1$, and 
the four-vectors of energy-momentum intensives ($g^a_1$, $g^a_2$,
dotted arrows) are light-like. }
\end{figure}

\begin{enumerate}
\item $w_2=0$:   the current stands in the observed body. 
In this case $w_1=v$, the measured energy current  speed is 
that of the moving body,
and \hbox{$T_1=T_2\sqrt{1-v^2}<T_2$},  the moving body appears cooler by a
Lorentz factor \cite{Pla07a,Pla08a,Ein07a} (see Fig \ref{Fig2}).

\item {$w_1=0$}: the current stands in the thermometer. 
In this case we must have $w_2=-v$ and \hbox{$T_1=T_2/\sqrt{1-v^2} > T_2$}, 
the moving body appears hotter {\cite{Bla47a,Ott63a,TerWer71a,Req08m}} (see Fig \ref{Fig3}).

\item $w_1+w_2=0$: the current is standing in the total system of moving
body and thermometer, the individual contributions exactly compensate
each other. This is achieved by a special value
of the energy current velocities, $w_2=-w$, $w_1=w$ with
$ w = (1-\sqrt{1-v^2})/{v}$.
In this case even the apparent temperatures are equal, $T_1=T_2$ \cite{Lan66a,Lan67a}
(see Fig \ref{Fig4}).

\item $w_2=1$: a radiating body (e.g. a photon gas) is moving. In
this case $w_1=1$, and one obtains $T_1=T_2\sqrt{\frac{1-v}{1+v}}$.
It means that $T_1 < T_2$ for $v>0$, a Doppler red shifted temperature
is measured for an aparting body (see Fig \ref{Fig5}) - quite common for astronomical objects
- and $T_1 > T_2$ for $v < 0$, a Doppler blue shifted temperature
appears for closing bodies - more common in high energy accelerator experiments.
\end{enumerate}

On figures \re{Fig2}-\re{Fig5} we fix the reference frame to the thermometer,
therefore $u_1^a= (1,0)$ (the vertical axis is time). The energy current
velocity four-vectors are perpendicular to the corresponding four-velocities,
therefore they are on lines symmetrical to the light cones. The four-velocity
vectors end on the timelike hyperbolas and the spacelike energy current velocities
end inside the spacelike hyperbolas. The temperature ratios are determined
by the magnitudes of the $g^a$-s  as $T_1/T_2= \|g_1^a\|/\|g_2^a\|$ (see
\cite{VanBir09dem}.

\section{Lorentz scalar temperature} According to the classical ansatz \(g^a
= \bar u^a\), the total entropy has to depend on the total energy 
$E=\ave_a{E}^a$. Then the equilibrium conditions \re{VEL_EQUIL_ECKART}
and \re{ECKART_TEMP} result in zero relative velocity $v=0$ and the temperatures
are equal  $T_1=T_2$.

However, thermodynamic and generic stability considerations
are favoring an other Lorentz-scalar combination $\|{E}\| =
\sqrt{{E}_a{E}^a}$ \cite{VanBir08a,BirAta08a,Van08a,Van09a}. Denoting the partial 
derivative of entropy $S(\|E\|,V)$ with respect
to its first argument by $1/\theta = \pd{S}{\|E\|}$, one re-writes the total
differential,
\be
 dS = \frac{1}{\theta}d\|E\| + \frac{\tilde{p}}{\theta} dV = \frac{g_a}{T} dE^a + \frac{p}{T}dV,
\ee{SCALAR_EOS_DIFF}
and comparing to the general Gibbs relation \re{GIBBS_NEW_FORM} one  obtains the correspondence
\be
 \frac{g^a}{T} = \frac{1}{\theta} \frac{E^a}{\|E\|}, 
 \qquad  \frac{p}{T} = \frac{\tilde{p}}{\theta}.
\ee{SCALAR_INTENSIVES}
It follows that the length of the intensive four-vector, $g^a$, is the ratio of
the traditional (energy associated) and scalar (energy-momentum four-vector length
associated) temperatures: $\sqrt{g_ag^a}=T/\theta$. On the other hand its projection to the average velocity
reveals the value in the comoving system:
\be
 g_a\ave^a = \frac{T}{\theta} \, \frac{E_a\ave^a}{\|E\|} = 1.
\ee{SCALAR_INTENSE_CO_MOVING}

This equation relates the energy-momentum-associated scalar temperature, $\theta$ 
to the energy-associated one, $T$. As a consequence we obtain 
\be
 {g^a} = \frac{E^a}{E^b \bar u_b}, 
 \qquad   {w^a} = \frac{E^a- \bar u^a (E^b \bar u_b)}{E^c \bar u_c}.
\ee{INTERPRETATION} 

The later formula  clearly interpret $w^a$ as the quotient of the comoving,
average velocity related  energy current (momentum) and energy of the thermodynamic
body, that is the energy current velocity. 

Finally we remark, that in the simple  two dimensional particular case we obtain that {$\theta=T/\sqrt{1-w^2}$}.
Therefore the equilibrium condition \re{SCALAR_TEMP_EQUIL} gives equal scalar 
temperatures: $\theta_1=\theta_2$. This is a stronger reflection of Landsberg's view,
and his physical arguments in \cite{Lan66a,Lan67a} than the assumption of zero total 
energy current velocity.

\section{Summary} We investigated the possible derivation of basic thermodynamical
laws for homogeneous bodies from relativistic hydrodynamics. 
The dependence of entropy on internal energy is replaced by a dependence on the
energy-momentum four-vector, $E^a$. 
As a novelty a relativistic heat four-vector has been formulated.
For the traditional, energy exchange related temperature, $T$, a 
universal transformation formula is obtained. 
For a general observer four 
velocities are involved in the equilibrium condition of two thermodynamic
bodies in equilibrium. One of them can be eliminated 
by choosing the observing frame, the physical relation depends only on the
relative velocity. Another condition connects the internal heat currents
in the bodies in thermal contact. So there remains two velocity like parameters
to describe thermal equilibrium: the energy current speed (the velocity
related to the integrated internal heat current density) in one of the bodies and
their relative velocity. The traditional temperature transformation formulas belong
to corresponding particular choices on the energy current speeds. This can
be the reason that no agreement could be achieved historically. For most
common cases there is no heat current in the observed body but it flows in the thermometer.
This leads to the Planck-Einstein transformation formula.

The closer relation to dissipative hydrodynamics favors a particular dependence of entropy 
on energy-momentum and leads to a Lorentz scalar temperature.

Our approach is covariant, and the compatibility to hydrodynamics clarifies 
that the Planck-Ott imbroglio is not a problem
of synchronization as it was supposed in \cite{Yue70a,DunHan09a}.
It makes possible to interpret the classical paradoxical results of Planck
and Einstein, Ott,  Landsberg and  Doppler in a   unified treatment. Our investigations reveal that despite
of the  apparent 
paradoxes related to Lorentz transformations, there is a covariant   
relativistic thermodynamics  with proper absolute temperature in full agreement 
with relativistic hydrodynamics.

\section{Acknowledgement}

The authors thank to L. Csernai for his enlighting remarks.

\bibliographystyle{unsrt}

\end{document}